\documentclass[twocolumn,showpacs,preprintnumbers,amsmath,amssymb,superscriptaddress]{revtex4-1}

\usepackage{graphicx}
\usepackage{dcolumn}
\usepackage{bm}

\begin{document}


\title{Selective probing of photo-induced charge and spin dynamics in the bulk and surface of a topological insulator}

\author{D. Hsieh}
\affiliation{Department of Physics, Massachusetts Institute of Technology, Cambridge, MA 02139, USA}
\author{F. Mahmood}
\affiliation{Department of Physics, Massachusetts Institute of Technology, Cambridge, MA 02139, USA}
\author{J. W. McIver}
\affiliation{Department of Physics, Massachusetts Institute of Technology, Cambridge, MA 02139, USA}
\affiliation{Department of Physics, Harvard University, Cambridge, MA 02138, USA}
\author{D. R. Gardner}
\affiliation{Department of Physics, Massachusetts Institute of Technology, Cambridge, MA 02139, USA}
\author{Y. S. Lee}
\affiliation{Department of Physics, Massachusetts Institute of Technology, Cambridge, MA 02139, USA}
\author{N. Gedik}
\affiliation{Department of Physics, Massachusetts Institute of Technology, Cambridge, MA 02139, USA}

\date{\today}

\begin{abstract}
Topological insulators possess completely different spin-orbit coupled bulk and surface electronic spectra that are each predicted to exhibit exotic responses to light. Here we report time-resolved fundamental and second harmonic optical pump-probe measurements on the topological insulator Bi$_2$Se$_3$ to independently measure its photo-induced charge and spin dynamics with bulk and surface selectivity. Our results show that a transient net spin density can be optically induced in both the bulk and surface, which may drive spin transport in topological insulators. By utilizing a novel rotational anisotropy analysis we are able to separately resolve the spin de-polarization, intraband cooling and interband recombination processes following photo-excitation, which reveal that spin and charge degrees of freedom relax on very different time scales owing to strong spin-orbit coupling.
\end{abstract}

\maketitle

Three-dimensional topological insulators \cite{Moore_Nature,RMP,Qi} are a promising new platform for spin-based electronics because of their unique spin-orbit coupled electronic structure, which is spin-degenerate and fully gapped in the bulk yet spin-polarized and gapless on all surfaces \cite{Hsieh_Nature,Hsieh_Science}. Excitation with light is predicted to drive novel bulk and surface responses including quantized magneto-optical rotation \cite{Qi_QFT}, bulk topological quantum phase transitions \cite{Inoue,Linder} and surface spin transport \cite{Raghu,Lu,Hosur}. Therefore understanding photo-induced charge and spin dynamics of both bulk and surface states in real materials is imperative for device applications. However knowledge of such out-of-equilibrium processes is lacking because conventional dynamical probes such as transport \cite{Hadar,Transport} or optics \cite{Sushkov,LaForge} measure a steady state response that is integrated over both the surface and bulk states.

\begin{figure}[h]
\includegraphics[scale=0.81,clip=true, viewport=0.0in 0in 6.8in 3.2in]{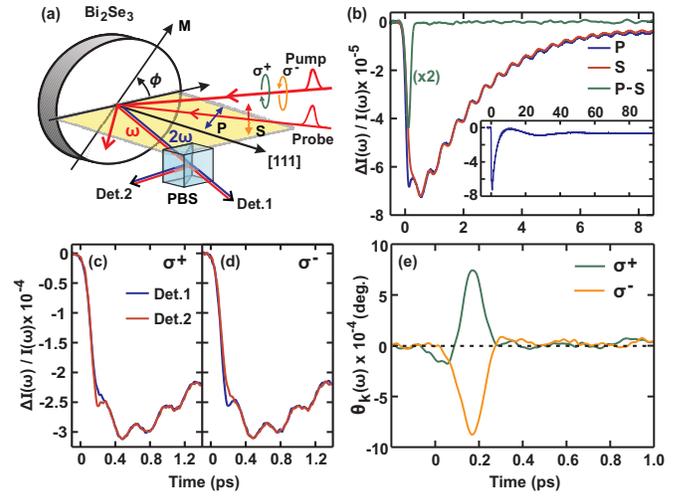}
\caption{\label{fig:Fig1} (a) Schematic of the experimental geometry. M denotes a mirror plane of the Bi$_2$Se$_3$ crystal. When the polarizing beam splitter (PBS) is oriented as drawn, the $s$- and $p$-polarized output probe photons are measured using detectors 1 and 2 respectively. The PBS is rotated by 45$^{\circ}$ to perform a balanced detection measurement of Kerr rotation. (b) The change in reflectivity of $s$ and $p$ fundamental probe photons following a $p$ pump pulse. The difference of the two traces is shown magnified by a factor of 2. Inset shows the $s$-out trace measured out to long delay times. (c) Typical pump-probe signal measured from detectors 1 and 2 with the PBS rotated by 45$^{\circ}$ following excitation by obliquely incident left ($\sigma^+$) and (d) right ($\sigma^-$) circularly polarized pump pulses. (e) $\sigma^+$ and $\sigma^-$ pump induced Kerr rotation $\theta_K$ obtained via the signal difference of detectors 1 and 2 from (c) and (d).}
\end{figure}

In this Letter, we selectively study the dynamic charge and spin photo-responses of both bulk and surface carriers in a prototypical topological insulator Bi$_2$Se$_3$ \cite{Xia,Zhang,Hsieh_Nature2,ARPES} using time-resolved spectroscopies. Our experiments utilize a pump-probe scheme where an ultrafast laser (pump) pulse excites a non-equilibrium charge or spin distribution in the material and a time-delayed (probe) pulse tracks the relaxation of the excited charge or spin population either through its time-resolved reflectivity \cite{Gedik_PRB} or Kerr rotation \cite{Kimel} respectively. The bulk and surface responses can be separately measured by exploiting the nonlinear optical properties of Bi$_2$Se$_3$ as follows. In general, the electrical polarization $P_i$ of any material contains frequency components at all harmonics of the driving field $E_j(\omega)$, where $\omega$ is the optical frequency and the indices run through three spatial coordinates. The fundamental response $P_i(\omega)=\chi_{ij}^{(1)}E_j(\omega)$ is given by a second rank susceptibility tensor $\chi_{ij}^{(1)}$ whose non-vanishing elements are determined by the crystal symmetry. Because non-zero $\chi_{ij}^{(1)}$ elements are allowed under the bulk crystal symmetry constraints of Bi$_2$Se$_3$ (space group $D^5_{3d}$ \cite{Zhang}), it is well known that the fundamental response originates predominantly from bulk carriers \cite{Sushkov,LaForge}. On the other hand, it is expected and experimentally shown \cite{Hsieh_SHG} that second harmonic generation (SHG) has two contributions $P_i(2\omega)=\chi_{ijk}^{(2)}E_j(\omega)E_k(\omega)+\chi_{ijkl}^{(3)}\varepsilon_j(0)E_k(\omega)E_l(\omega)$ that vanish everywhere except at the surface. The third rank tensor $\chi_{ijk}^{(2)}$ vanishes in any inversion symmetric crystal such as Bi$_2$Se$_3$ and is only allowed at the surface where inversion symmetry is necessarily broken. Although the fourth rank tensor $\chi_{ijkl}^{(3)}$ can be non-zero in the bulk, the static depletion electric field $\varepsilon_{j}(0)$ is only non-zero within a thin ($\sim$ 2 nm) space-charge region near the surface \cite{Analytis}.

Fundamental and second harmonic reflections were measured from the (111) surface of lightly arsenic doped metallic Bi$_2$Se$_3$ bulk single crystals \cite{Hadar}, which were cleaved in air at room temperature. Experiments were performed using 795 nm (1.56 eV), 80 fs laser pulses from a Ti:sapphire oscillator with a 1.6 MHz repetition rate. The weaker probe pulse was derived from the pump pulse with a beam splitter and then passed through a delay line. The typical pump fluence was $\sim$0.4mJ/cm$^2$. The incidence angle of the pump was varied between 0$^{\circ}$ (normal) and 60$^{\circ}$ (oblique) while the incidence angle of the probe was fixed around 45$^{\circ}$. Both beams were always kept in the same scattering plane and were focused and spatially overlapped onto a 20 $\mu$m spot on the sample. Reflected probe photons at $\omega$ were detected using Si photodiodes and reflected 2$\omega$ photons, which were spectrally isolated through a combination of interference and absorptive filtering, were detected using calibrated photomultiplier tubes. Polarization rotation was measured using a balanced detection scheme [Fig.\ref{fig:Fig1}(a)].

We first investigate the transient bulk response to photo-excitation by measuring the fundamental pump-probe signal. Fig.\ref{fig:Fig1}(b) shows the temporal change in the reflected intensity $\Delta I(\omega)/I(\omega)$ of both linearly $s$- and $p$-polarized probe light following excitation by a $p$-polarized pump pulse, which generates a non-equilibrium charge distribution in the material via interband transitions since the incident photon energy far exceeds the bulk gap ($\sim$0.3 eV). Traces taken with both probe polarizations are identical at all times ($t$) except during the pump-probe overlap time $0<t<160$ fs, which originates from a coherent interference between $p$ pump and $p$ probe beams \cite{Vardeny}. Following the fast initial dip, the reflectivity undergoes a slow recovery that can be described by an exponential with a time constant of $\sim$ 2.3 ps \cite{EPAPS}. A similar decay has also been observed among the same materials family \cite{Qi_ultrafast,Wu} and is attributed to the cooling of photo-excited carriers through electron-phonon scattering. The fast oscillatory component with frequency 2.16 THz that is superimposed on the decay can be attributed to a pump induced coherent vibration of the A$_{1g}$ longitudinal optical phonon, and the low frequency oscillation [Fig.\ref{fig:Fig1}(b) inset] to coherent longitudinal acoustic phonons \cite{Qi_ultrafast}.

To investigate the transient bulk spin response of Bi$_2$Se$_3$ we excite the material with circularly polarized pump pulses, which is known to generate a non-equilibrium spin polarized charge distribution in semiconductors such as GaAs and Si owing to the optical orientation effect \cite{Meier}. Transient magnetization ($\vec{M}$) can be measured through the Kerr effect, which is a rotation of the polarization plane of probe light with wave vector $\vec{k}$ by an angle $\theta_K(\omega) \propto \vec{M}\cdot\vec{k}$ \cite{Kimel}. A finite $\theta_K(\omega)$ is manifested as an intensity difference between two detectors in balanced detection geometry [Fig.\ref{fig:Fig1}(a)]. Fig.\ref{fig:Fig1}(c) and (d) display typical traces measured in two such detectors using obliquely incident circular pump and $p$ probe light. These traces indicate that a Kerr rotation [Fig.\ref{fig:Fig1}(e)] exists when the pump and probe are temporally overlapped and changes sign depending on whether the pump light is left ($\sigma^+$) or right ($\sigma^-$) circularly polarized. This phenomenon cannot be explained by coherent interference because the $p$ probe polarization is an equal superposition of $\sigma^+$ and $\sigma^-$. Rather, it is naturally explained by optical spin orientation via the inverse Faraday effect where circularly polarized pulses induce a helicity dependent magnetization $\vec{M}$ $\propto$ $\vec{E}(\omega)$$\times$$\vec{E}^{*}(\omega)$ \cite{Kimel}. The rapid decay of $\theta_K(\omega)$ is consistent with strong spin-orbit coupling in Bi$_2$Se$_3$, which is expected to de-polarize any transient $\vec{M}$ within a mean free time $\sim$100 fs \cite{Analytis}.

\begin{figure}
\includegraphics[scale=1.09,clip=true, viewport=0.0in 0in 6.8in 2.7in]{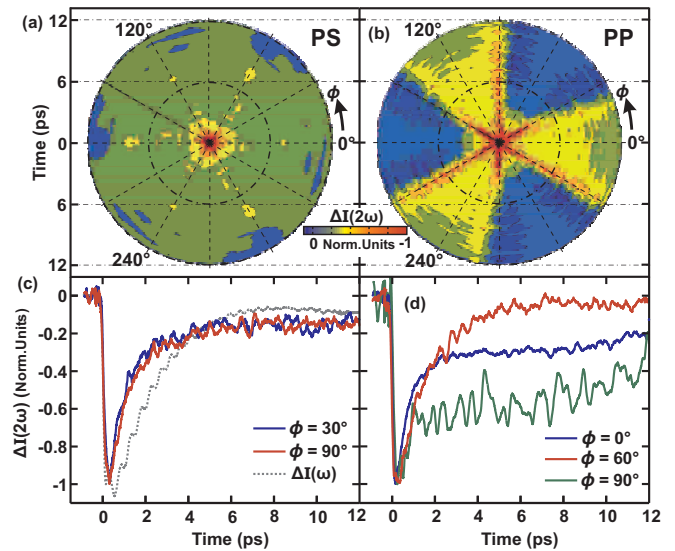}
\caption{\label{fig:Fig2} (a) $p$ pump induced change in SHG intensity as a function of sample angle $\phi$ measured with $p$-in $s$-out and (b) $p$-in $p$-out probe photons. Data are normalized to their minimum. Data are taken in the range $0^{\circ}<\phi<120^{\circ}$ and then three-fold symmetrized. (c) and (d) show constant $\phi$ cuts through the data in (a) and (b) respectively. A normalized fundamental pump probe trace taken with the same pump fluence is overlayed in (c).}
\end{figure}

Having understood the bulk response of Bi$_2$Se$_3$ to charge and spin excitations, we proceed to study the corresponding surface responses using SHG. Unlike the fundamental intensity, the SHG intensity $I$($2\omega$) from Bi$_2$Se$_3$ depends on the orientation $\phi$ [Fig.\ref{fig:Fig1}(a)] of the light scattering plane relative to the crystal mirror plane through the relations \cite{Hsieh_SHG}:

\begin{eqnarray}
I_{ps}(2\omega) &=& |a\sin3\phi|^2 \\ \vspace{0.5cm}
I_{pp}(2\omega) &=& |b+a\cos3\phi|^2 \nonumber
\end{eqnarray}

\noindent where subscripts on the intensity denote the input and output polarization of the probe beam and $a$ and $b$ are linear combinations of $\chi^{(2)}$ and $\chi^{(3)}\varepsilon(0)$ tensor elements that describe the in- and out-of-plane components of the surface response respectively \cite{Hsieh_SHG}. In order to selectively investigate both components of the transient surface response to charge excitation, we measure the $p$ pump induced change in SHG intensities $\Delta I_{ps}(2\omega)$ and $\Delta I_{pp}(2\omega)$ over the complete range of $\phi$ and normalize each trace by its minimum value. Fig.\ref{fig:Fig2}(a) shows that the normalized $\Delta I_{ps}(2\omega)$ traces are isotropic in $\phi$, which is consistent with being exclusively sensitive to the transient in-plane response $|a|^2$ in Eq.1. The stronger intensity fluctuations observed along the mirror planes ($\phi = 0^{\circ}, 60^{\circ}, 120^{\circ}$) is an artifact of the vanishing SHG intensity at these angles (Eq.1). Individual pump-probe traces [Fig.\ref{fig:Fig2}(c)] can all be fit to an exponential decay with a time constant of $\sim$ 1.2 ps \cite{EPAPS}, which shows that surface carriers excited across the bulk energy gap are cooled faster than those in the bulk, possibly due to enhanced surface electron-phonon scattering or diffusion of hot surface carriers into the bulk. 

Conversely, the normalized $\Delta I_{pp}(2\omega)$ is strongly anisotropic in $\phi$ and exhibits the three-fold rotational symmetry and mirror symmetries of the surface crystal structure [Fig.\ref{fig:Fig2}(b)]. The slight differences between the $\pm30^{\circ}$ traces are again artifacts of the large fluctuations associated with a weak SHG intensity at these angles. Traces at $\phi = 0^{\circ}$ and $60^{\circ}$ [Fig.\ref{fig:Fig2}(d)], which measure the time dependence of $|b+a|^2$ and $|b-a|^2$ respectively (Eq.1), exhibit an initial fast decay similar to $\Delta I_{ps}(2\omega)$ for $t < 2$ ps but reach values greater or less than $\Delta I_{ps}(2\omega)$ for $t > 2$ ps depending on the relative sign of $a$ and $b$. This indicates that the transient change in the out-of-plane polarization $b$ is significantly smaller than that of the in-plane polarization $a$ and decays at a much slower rate. The time dependence of $|b|^2$ can be selectively studied at $\phi = 90^{\circ}$ where the contribution from $a$ vanishes according to Eq.1. Fig.\ref{fig:Fig2}(d) shows that $|b|^2$ follows a slow exponential decay with a time constant of $\sim$ 21 ps \cite{EPAPS}, which can be attributed to changes in $\varepsilon(0)$ as follows. A space-charge layer near the surface is known to exist in Bi$_2$Se$_3$ owing to a migration of negatively charged Se vacancies to the surface \cite{Hsieh_Nature2,Hsieh_SHG}. This creates an internal out-of-plane electric field $\varepsilon(0)$ that penetrates over a screening length determined by the density of free carriers and should only affect the out-of-plane electrical polarization $b$. It has been experimentally shown that $I(2\omega)$ increases with increasing $\varepsilon(0)$ \cite{Hsieh_SHG}. Because interband photo-excitation generates additional free carriers that can act to screen $\varepsilon(0)$ \cite{Chang}, this leads to a negative contribution to $\Delta I(2\omega)$. Therefore the time dependence of $|b|^2$ must represent the recovery of $\varepsilon(0)$ via electron-hole recombination of bulk-like states at the surface across the bulk gap, which is expectedly slower than the intraband cooling $|a|^2$.

\begin{figure}[h]
\includegraphics[scale=0.85,clip=true, viewport=0in 0in 6.8in 3.4in]{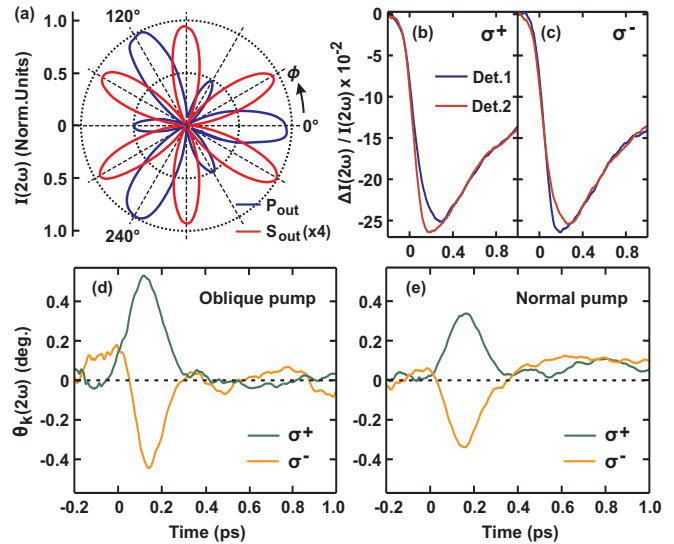}
\caption{\label{fig:Fig3} (a) Static SHG intensity as a function of $\phi$ measured using $p$-in and $p$-out or $s$-out probe photons. Time-resolved SHG Kerr rotation measurements are taken with $p$ pump and $\phi$ = 0$^{\circ}$ so that $\theta_K$ = 0$^{\circ}$ when there is no pump. (b) Typical SHG pump-probe signal measured from detectors 1 and 2 with a 45$^{\circ}$ oriented analyzer following excitation by obliquely incident $\sigma^+$ and (c) $\sigma^-$ pump pulses. (d) $\sigma^+$ and $\sigma^-$ pump induced Kerr rotation measured at oblique pump incidence and (e) normal pump incidence. We note that there is some statistical fluctuation in the temporal widths of the Kerr rotation peaks due to noise in the raw data traces (b) and (c).}
\end{figure}

Having established a two-step surface charge relaxation process following charge excitation, we investigate the surface response to spin excitations using circularly polarized pump pulses. Unlike reflected fundamental light, reflected SHG light is allowed to be rotated with respect to the linearly polarized incident probe beam even in the absence of a pump pulse due to the off diagonal elements of $\chi^{(2)}$ and $\chi^{(3)}$ \cite{Hsieh_SHG}. Fig.\ref{fig:Fig3}(a) shows that incident $p$ probe light will generate SHG from Bi$_2$Se$_3$ with an $s$ component except when $\phi$ coincides with a crystal mirror plane. To avoid this intrinsic optical rotation and to maximize sensitivity to pump-induced rotation, we perform spin-sensitive time-resolved measurements at $\phi = 0^{\circ}$. Fig.\ref{fig:Fig3}(b) and (c) display typical traces measured in two detectors in balanced detection geometry using obliquely incident circular pump and $p$ probe pulses, which, like the fundamental response [Fig.\ref{fig:Fig1}(c)\&(d)], show evidence for a pump helicity dependent optical rotation only during the pump-probe overlap time. There are two possible microscopic mechanisms for this effect. One is an inverse Faraday effect from bulk-like bands \cite{Xia,Hsieh_Nature2,ARPES} near the surface and the other is a photo-induced magnetization of the Dirac surface states of a topological insulator. The latter effect arises because topological Dirac surface states have a helical spin texture \cite{Hsieh_Science,Hsieh_Nature2} where spins are polarized in the surface plane perpendicular to their momentum and rotate by 2$\pi$ around the Dirac cone such that there is no net magnetization in equilibrium \cite{Hsieh_Science,Hsieh_Nature2}. However because an obliquely incident circular photon pulse can excite spins asymmetrically in $\vec{k}$-space owing to angular momentum selection rules, the Dirac cone can acquire a net out-of-equilibrium magnetization \cite{Hosur,Lu,Raghu}. Such an effect has already been observed in Bi$_2$Se$_3$ using angle-resolved photoemission spectroscopy with 6 eV light \cite{Wang}. To address whether a magnetized Dirac cone or a surface inverse Faraday effect is dominant using 1.5 eV light, we note that a normally incident pump pulse couples uniformly to all planar spins around the Dirac cone and therefore cannot create a net magnetization. Although higher order effects can introduce some $\vec{k}$ dependent out-of-plane canting of the spins \cite{Fu_warp} and thus allow non-uniform excitation by normally incident light, these canting angles are known to be very small. On the other hand, the inverse Faraday effect will generate a non-zero Kerr rotation even for a normally incident pump because $\vec{M}\cdot\vec{k}\neq0$ as long as the pump and probe wave vectors are not orthogonal. In our experimental geometry, to a linear approximation, we may expect the ratio of $\theta_{K}(2\omega)$ between normal and oblique pump incidence geometries to be $\cos(45^{\circ})/\cos(15^{\circ})\sim0.73$. Our normal incidence measurement [Fig.\ref{fig:Fig3}(e)] shows that $\theta_K(2\omega)$ remains non-zero and reaches a magnitude $\sim$0.7 times the oblique case, which suggests that it is an inverse Faraday effect of bulk-like bands at the surface and not a magnetization of the Dirac cone that is primarily responsible for the surface optical rotation we observe.

\begin{figure}
\includegraphics[scale=0.34,clip=true, viewport=0.0in 0in 10.0in 5.4in]{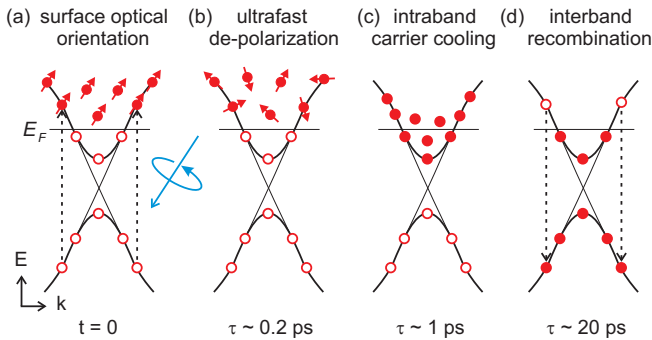}
\caption{\label{fig:Fig4} Schematic of the transient surface charge and spin response to photo-excitation and their characteristic time scales $\tau$. Parabolic curves represent the valence and conduction bands of Bi$_2$Se$_3$ near the surface and the straight lines traversing the gap represent the Dirac surface states. Empty circles represent holes and filled circles represent electrons. (a) The circular pump pulse initially creates a spin-polarized excited carrier population. (b) The spin-polarization is then rapidly de-polarized because of spin-orbit coupling within the pump-probe overlap time. (c) Intraband carrier cooling occurs on a 1 ps time scale followed by (d) a much slower interband electron-hole recombination.}
\end{figure}

Our work shows that ultrafast circularly polarized pulses are able to generate a transient spin-polarized charge population in both the bulk states and bulk-like surface states of Bi$_2$Se$_3$ owing to the inverse Faraday effect and that spin and charge degrees of freedom relax on very different time scales following photo-excitation. For the surface in particular, we identified a three-step response, schematically illustrated in Fig.\ref{fig:Fig4}, that consists of a rapid spin de-polarization followed by intraband cooling via surface electron-phonon scattering and finally a much slower interband electron-hole recombination. Existing theoretical works about photo-induced spin transport on topological insulator surfaces only account for carriers in the Dirac cone \cite{Raghu,Lu,Hosur}. Our work shows that carriers photo-excited into bulk spin-degenerate bands also carry a net magnetization and may acquire a net drift velocity if, for instance, they are excited asymmetrically in momentum space from the Dirac cone or by accounting for the finite linear momentum transfer from photon absorption \cite{Grinberg}. More generally, we have demonstrated a method to selectively probe the charge and spin responses of both the bulk and surface states, and have developed a procedure using rotational anisotropy time-resolved SHG to selectively probe the in-plane and out-of-plane surface polarization responses. Performing these measurements at photon energies below the bulk gap may reveal signatures of the Dirac cone response at both a bare surface or buried solid interface.

We thank Darius Torchinsky, Yihua Wang and Stefan Kehrein for helpful discussions. This work is supported by D.O.E. Grant No. DE-FG02-08ER46521

\end{document}